# Interconnected Markets: Exploring the Dynamic Relationship Between BRICS Stock Markets and Cryptocurrency

Wei Wang

College of Economics and Management, Chang'an University, Xi'an, Shaanxi, 710064, China

wwang@chd.edu.cn

Haibo Wang

*Department of International Business and Technology Studies, Texas A&M International University, Laredo, Texas, USA*

hwang@tamui.edu

Wendy Hong Wang

Department of Computer Science and Information Systems, University of North Alabama, Florence, AL 35632, USA

hwang21@una.edu

Martin Enilov

Southampton Business School, University of Southampton

12 University Rd Southampton, SO17 1BJ

Email: m.p.enilov@soton.ac.uk

Corresponding authors: Wendy Hong Wang, hwang21@una.edu; Haibo Wang, hwang@tamiu.edu

**Abstract**: This study aims to examine the intricate dynamics between BRICS traditional stock assets and the evolving landscape of cryptocurrencies. Using a time-varying parameter vector autoregression model (TVP-VAR), we have analyzed data from the BRICS stock market index, cryptocurrencies, and indicators from January 6, 2015, to June 29, 2023. The results show that three out of the five BRICS stock markets serve as primary sources of shocks that subsequently affect the financial network. The transcontinental (TCI) value derived from the dynamic conditional connectedness using the TVP-VAR model demonstrates a higher explanatory power than the static connectedness observed using the standard VAR model. The discoveries from this study offer valuable insights for corporations, investors, and regulators concerning systematic risk and investment strategies.

**Keywords:** BRICS, Emerging Economies, Stock Market Linkages

## 1. Introduction

The globalization of stock markets is pivotal in the trajectory of international finance. The ongoing surge in the opening of stock markets worldwide propels global economic cooperation and trade exchanges in an escalating frequency (Wen et al., 2019). Notably, the 1990s was marked by the unfurling of this transformative process, as both developed and emerging market countries actively participated in the liberalization of their stock markets. Consequently, a convergence in the economic fundamentals among diverse nations has ensued, catalyzed by heightened collaborative economic endeavors. This convergence has deepened the interconnection between stock markets worldwide, highlighting the dynamic and interdependent nature of the international financial landscape (Uddin et al., 2021).

As global financial markets are increasingly interconnected, uncertainty shocks from external economies can significantly impact domestic financial markets, particularly those of emerging economies (Asafo-Adjei et al., 2021; Carrière-Swallow & Céspedes, 2013). As quintessential representatives of emerging economies and developing nations, the BRICS countries (Brazil, Russia, India, China, and South Africa) wield substantial economic influence (Khalfaoui et al., 2023) with a combined gross domestic product of 24.5 trillion US dollars in 2021, about one-quarter of the global total. As the BRICS countries advance open economy initiatives, scholarly attention has been increasingly drawn to the evolving stock market linkages within this bloc (Goodell et al., 2023). Given the inherent fragility of emerging economies and the intricate web of trade, capital flows, debt, and financial aid interdependencies, the monetary policies of major open economies, including the Eurozone, the United States (US), and China, have profoundly affected various financial indicators within the BRICS countries.

Since its inception in 2009, the digital currency market, epitomized by Bitcoin, has proliferated, garnering significant attention worldwide (Feng et al., 2018). As the digital currency market solidifies its position within the broader financial system, a noteworthy scenario emerges in times of crisis: many investors may redirect their assets to the digital currency market, potentially triggering the collapse of traditional financial markets (Bommer et al., 2023). Should the digital money market remain intricately linked with traditional financial markets, such as the stock market, the fallout from a collapse in the latter could reverberate through the former, intensifying systemic risk and resulting in more substantial losses (Rehman et al., 2023). Given the proliferation of digital currencies, this article aims to delve into the intricate interdependence between traditional stock assets and the evolving landscape of cryptocurrencies. The interconnections among global financial markets are tightening, marked by the emergence of a new international financial cycle that elevates the significance of spillover effects between countries or regions (Naeem et al., 2023). This study specifically investigates the interplay between the stock markets in BRICS countries and the realm of cryptocurrencies. We employ a time-varying parameter vector autoregression (TVP-VAR) model, delving into a comprehensive analysis of the spillover effects stemming from financial markets in BRICS nations and their interaction with the dynamic landscape of cryptocurrencies.

The rest of this paper is organized as follows: section 2 provides a comprehensive review of the existing literature, section 3 introduces the TVP-VAR framework and provides insight into the data compilation process, section 4 discusses the results of this study, section 5 summarizes the main findings, and section 6 concludes with pivotal insights gained from this study.

## 2. Literature review

### 2.1. Stock market dynamic connectedness

The dynamics of stock market interconnectedness reveal a continuous process where changes in one market prompt corresponding changes in another. This interconnectivity stems from factors like shared fundamentals, similar macroeconomic conditions, and stages of economic development. International fund flows, information exchange, and close economic ties further contribute to dynamic linkages between markets globally and within sectors and individual stocks.

As economic and financial globalization advances, collaboration among countries intensifies. The liberalization of international goods, services, technology, and capital flows propels global economic integration (Al-Mohamad et al., 2020; Pan et al., 2022). The internationalization of stock markets becomes crucial in this context, drawing the attention of cross-border investors (Cui et al., 2022; Younis et al., 2021); thus, stock market linkage analysis and factors influencing such linkages have become a central focus in international financial research (Papadamou et al., 2021; Zhang et al., 2020). In this research, we explore linkage relationships and causal relationships within the stock market. The international economic and trade transmission channel is crucial in stock market interconnectedness. Robust trade relations allow one nation's financial changes to affect others. With emerging economies rising, scholars broaden their perspectives to include the interdependence between developed and emerging market stock markets.

The US stock market has a guiding effect on emerging economies in the Asia Pacific. Analyzing stock market linkages within the BRICS nations becomes pivotal for mitigating the impact of regional or global economic crises, providing theoretical insights for stock market integration and the BRICS cooperation mechanism (Khan et al., 2020; Singh et al., 2022). Table 1 summarizes studies on the dynamic connectedness within stock markets.

**Insert Table 1 Here**

Research on linkages within the BRICS consortium often employs various models. This study takes a novel approach to utilizing the TVP-VAR model to investigate linkages between BRICS stock markets and cryptocurrencies. This unique approach has profound practical significance.

### 2.2. Cryptocurrency & financial assets

In recent years, cryptocurrency has become a decentralized alternative to fiat money, operating independently of sovereign governments and traditional banking systems. It has been recognized for addressing inefficiencies in conventional economic structures, attracting increasing attention from investors, practitioners, and researchers (Goodell et al., 2023). Investigating its potential impact on the stock market as an exogenous economic variable offers a novel research avenue.

Since the inception of digital currency such as Bitcoin, it has demonstrated its transformative impact on monetary systems, national economies, social governance, and international exchanges (Fofack et al., 2020; Yu et al., 2021), as well as substantial risks evident in Bitcoin's price fluctuations.

Amid waning confidence in traditional financial markets, investors turn to digital currencies as a "safe haven." Empirical evidence suggests that digital currencies can support traditional markets during market crises and provide alternative investment avenues. Some consider it a short-term hedging tool (Bouri et al., 2017), while others see it as a diversified transaction asset (Stensås et al., 2019). The cryptocurrency market's fluctuations and increased trading volumes have drawn attention, positioning it as an exchange medium and a new investment asset category due to its high volatility, fat tail characteristics, and leverage effects.

Nonetheless, the anonymity of cryptocurrency makes it susceptible to asset price bubbles, posing risks to financial system stability. Table 2 comprehensively reviews recent studies examining the relationship between cryptocurrency and financial assets. Exploring the correlation between cryptocurrency and traditional financial markets is imperative for identifying diversified investment opportunities, evaluating hedging strategies, and managing potential spillover effects.

**Insert Table 2 Here**

### 2.3. The Relationship between Cryptocurrency Risk and Stock Market Risk

Prior literature has utilized the risk spillover index method to examine risk transmission across various markets (Testa, 2021); this approach is based on the generalized prediction error variance decomposition proposed by Diebold et al. (2009). Scholars have employed this method to investigate risk spillover effects between the Bitcoin market and traditional stock markets, revealing bidirectional net spillover relationships, especially after introducing Bitcoin derivatives (Zhang et al., 2021). Periodic variations in these relationships indicate an ongoing trend toward integration between the Bitcoin and stock markets. While acknowledging risk transmission between these markets, the periodic variability underscores the Bitcoin market's susceptibility to external events and policy influences.

## 3. Data and methodology

### 3.1. Data description

Data collected for this study consists of the BRICS stock market index, cryptocurrency data, and investor sentiment indicators from January 6, 2015, to June 29, 2023. The BRICS stock market and cryptocurrency data were from the Center for Research in Security Prices (CRSP). Table 3 presents detailed descriptions of this data. We partition the data into two distinct samples: the period preceding the COVID-19 outbreak (January 6, 2015 to February 19, 2020), and the period following the outbreak (February 20, 2020 to June 29, 2023).

**Insert Table 3 Here**

### 3.2. TVP-VAR method

The VAR model has been widely embraced for its econometric excellence; however, it has limitations and is susceptible to unstable model parameters. The evolution of VAR models has progressed from the structural vector autoregression model to the Bayesian vector autoregression model and further expanded to the spatial metric penal vector autoregression model. Eventually, the VAR model advanced into nonlinearity and time-varying parameters, as exemplified by the time-varying parameter vector autoregression (TVP-VAR) model. Primiceri (2005) initiated the nonlinear TVP-VAR model, Diebold and Yilmaz (2009) proposed and implemented its theoretical background and how the TVP-VAR model reports indicators that can be used to measure the overall correlation between variables, and later on, Nakajima (2011) enhanced it to scrutinize the dynamic and time-varying impact between stock market volatility and cryptocurrency. The TVP-VAR model offers distinctive advantages. First, it captures intricate nonlinear time-varying relationships among economic variables by incorporating time-varying coefficients and a variance-covariance matrix; second, the Bayesian estimation ensures timely adjustments of model parameters, sidestepping the need for arbitrary window size selection and minimizing sample discards. Furthermore, it remains resilient against excessive persistence and insensitivity to outliers. Therefore, we choose the TVP-VAR model in this study to examine the stability of short-term variable correlations and to evaluate whether long-term relationships conform to theoretical expectations.

The basic TVP-VAR model is:

$$x_t = \sum_{j=1}^{p} \emptyset_t x_{t-j} + w_t, w_t \sim Normal(0, S_t) \tag{1}$$

$$\emptyset_t = \emptyset_{t-1} + v_t,\ v_t \sim Normal(0, R_t) \tag{2}$$

where $x_t$ is a vector of variables of interest, and $p$ is the best lag length based on a Bayesian information criterion. Both $w_t$ and $v_t$ are vectors of the estimated error, while $S_t$ and $R_t$ are time-varying variance-covariance matrices. Finally, $\emptyset_t$ is a matrix for the estimated coefficients of each variable at time *t*.

The shock spillover from variable $i$ to variable $j$ can be measured by a variance decomposition matrix, $D_{ij}(h)$, with element $d_{ij}(h)$ as the $h$-step ahead of predicted error variance decompositions; $l_{ij}(h)$ is the standardized value of $d_{ij}(h)$. The "receiver" of spillover effect is:

$$Receiver_i\ (h) = 100\ \frac{\sum_{j=1, i\neq j}^{N} l_{ij}(h)}{\sum_{j=1}^{N} l_{ij}(h)}. \tag{4}$$

The "givers" of spillover effect is given as:

$$Giver_i(h) = 100\ \frac{\sum_{j=1, i\neq j}^{N} l_{ji}(h)}{\sum_{j=1}^{N} l_{ij}(h)}. \tag{5}$$

The total connectedness index of the spillover effect is given as:

$$TCI(h) = 100\ \frac{\sum_{i,j=1, i\neq j}^{N} l_{ij}(h)}{\sum_{i,j=1}^{N} d_{ij}(h)} \tag{6}$$

The net pairwise directional connectedness (NPDC) is given as:

$$NPDC_{i\leftarrow j} = l_{ij} - l_{ji} \tag{7}$$

The pairwise connectedness index of the spillover effect is decomposed from TCI and is given as:

$$PCI_{ij}(h) = \frac{l_{ij} + l_{ji}}{l_{ii} + l_{jj} + l_{ij} + l_{ji}} \tag{8}$$

and the pairwise influence index of the spillover effect is given as:

$$PII_{ij}(h) = \frac{l_{ij} - l_{ji}}{l_{ij} + l_{ji}} \tag{9}$$

The TVP-VAR model measures the dynamic conditional connectedness among variables by evaluating the average impact of a shock in each variable and its volatility on all the other variables. The $Receiver_i$ connectedness index represents the directional spillover received by variable $i$ from all other variables. The $Giver_i$ connectedness index represents the directional spillover sent by variable $i$ to all other variables. The differences between the total directional $Giver_i$ and $Receiver_i$ is given as $NET_i$, which is the net influence of variable $i$. "Net Pairwise Directional Connectedness (NPDC)" measures the influence/spillover variable $i$ has on variable $j$. If NPDC is positive, then $i$ dominates $j$; otherwise, $j$ dominates $i$. In general, the value of net spillover could suggest the position of investors in risk-hedging strategies. The negative value of $NET_i$ indicates the long position of the portfolio, and the positive value indicates the short position.

For an extensive network, $NPDC_{i,j}$ are computationally efficient and more accurate than $TCI$ in reducing bias results. The value and sign of the net total directional connectedness identify whether a specific variable is a net giver or receiver of uncertainty shocks over time and its rank in the network. The value and sign of $NPDC_{i,j}$ indicate the dynamic conditional connectedness between two specific variables. The $NPDC_{i,j}$ measures between variables are also graphically reported.

# 4. Empirical findings

## 4.1. Data preprocessing

Table 4 presents the descriptive statistics for stock markets in Brazil (EWZ), India (INDA), South Africa (EZA), China (MCHI), (Russia) ERUS, and the Volatility Index (VIX). The medians of all variables, except VIX, are positive and close to zero, with BTC exhibiting the highest value. BTC also displays the highest standard deviation among the seven portfolios, indicating greater volatility than other variables. All variables, except VIX, demonstrate negative skewness, signifying a longer left tail in their probability density function. In contrast, VIX exhibits positive skewness, indicating a longer right tail. MCHI has a skewness value close to zero, resembling the shape of a standard normal distribution. The elevated kurtosis values suggest fat tails in the distribution of these variables. The results of the Ljung-Box Q test reveal serial correlation in squared series for all variables, implying time-varying variance in each variable (Ljung & Box, 1978). Consequently, the TVP-VAR model is deemed suitable for assessing the connectedness network among the variables.

**Insert Table 4 Here**

In conjunction with descriptive statistics, we conduct an augmented Dickey-Fuller (ADF) unit-root test to assess the integration properties of the variables. The results demonstrate that all variables are stationary at the first-order difference across all samples (Table 5), suggesting that these variables are integrated into order one. Consequently, the first-order difference form of the variables is employed in subsequent analyses.

**Insert Table 5 Here**

We also conduct a CHOW test to assess the nonlinear relationship; the results suggest that compared to the traditional linear VAR model, the TVP-VAR model can capture the nuanced nonlinear relationships among variables; hence, a more practical approach to address the nonlinearity inherent in stochastic volatility.

### 4.2. Connectedness network

A series of tests for dynamic connectedness was employed to estimate spillover contributions and the impact of cryptocurrency and investor sentiments on the BRICS stock markets. Given the pairwise cointegration among the variables, a VAR connectedness model with time variation was executed for portfolio connectedness analysis. The results show that EZA, INDA, and EWZ emerge as the primary contributors of spillover shocks in the financial network, with EZA making the highest contribution and EWZ the lowest, whereas MCHI and ERUS are identified as the recipients of spillover shocks (See Table 5).

**Insert Table 6 Here**

MCHI, ERUS, VIX, and BTC are the recipients of spillover contributions on BRICS stock markets, with VIX experiencing the most significant influence from others. The diagonal elements in Table 6 signify the self-variable spillovers of financial markets, while the remaining elements depict the interactions across variables. The average total contribution to interconnectedness (TCI) in the static connected network is 51.82%, suggesting that the shock of all other variables can explain 51.82% of the forecast error variance of one variable. The "NET" row indicates that VIX appears to be the most significant net giver with a NET value of −10.81%. Consequently, VIX is likely to be influenced by other variables. In contrast, EZA emerges as the most substantial net giver in the network, boasting an average net connectedness value of 13.00%. Both EZA and INDA are likely to impact other variables. Figure 1 presents the graphical representation of static connectedness, where nodes in blue act as net givers, and nodes in yellow are net receivers of shocks in the financial network. Arrows in the graph indicate greater directional connectedness concerning net connectedness. Node sizes in the graph reflect the weighted average net total directional connectedness in the financial network. Bold lines in the graph denote a higher spillover than fine lines between variables. The bold line between EZA and VIX signifies a more significant spillover from EZA to VIX; similar bold lines are observed between other pairs of nodes, such as EZA and ERUS, EZA and MCHI, INDA and VIX, EWZ and EZA, INDA and ERUS. Thus, higher contributions are noted on portfolios with elevated ESG scores; however, no bold line is observed between BTC and other variables. One plausible explanation is that BTC might not have a linear relationship with other variables, and VAR connectedness analysis might not capture the nonlinear relationship.

The static connectedness measures from VAR offer an overview of the underlying interrelations in the financial network when variables exhibit linear relationships; however, the TVP-VAR model (See Table 7) can provide more nuanced insights into the time-varying connectedness among the variables, especially when nonlinear relationships are at play.

**Insert Figure 1 Here**

Figure 1 shows the NPDC measure plot of VAR connectedness with time. The TVP-VAR graphics visually depict the dynamic conditional connectedness levels among all pairs within the financial network. Nodes in blue represent net givers, while nodes in yellow represent net receivers of an uncertainty shock. Arrows in the graph signify greater directional connectedness in terms of time-varying pairwise net connectedness. The weighted time-varying net total directional connectedness determines node sizes. Bold lines indicate a higher spillover level than fine lines between variables over time.

Regarding dynamic conditional connectedness, we utilize the TCI to measure each variable's average impact on all others. The analysis highlights EZA, INDA, and MCHI as the primary contributors to shocks in the financial network, with EZA making the highest contribution and INDA the lowest. Conversely, EWZ, ERUS, VIX, and BTC emerge as net receivers of spillover shocks in the financial network, with BTC experiencing the most significant influence from others. Consequently, BTC exhibits a robust effect on critical mineral portfolio returns, and shocks in return spillovers and volatility influence investor behavior and carbon emissions forecasts.

**Insert Table 7 here**

The TCI value in Table 7 reveals that the average dynamic conditional connectedness in the financial network is 51.89%, signifying that the shock spillovers of all other variables can explain the forecast error variance of one variable within this network. Notably, the TCI for dynamic conditional connectedness surpasses the TCI for static connectedness from the VAR model. BTC emerges as the largest net receiver based on the NET value, boasting an average net connectedness value of −5.54%, implying that BTC is prone to receiving shocks from other variables. This negative net spillover also occurs in EWZ, ERUS, and VIX, with values of −1.49%, −4.04%, and −3.48%, respectively. In contrast, EZA is the largest net giver in the financial network, exhibiting a net connectedness value of 9.95%. The positive net spillover is evident in EZA, INDA, and MCHI, indicating that these variables are net givers of shocks. A notable shift from the static connectedness observed in the VAR model is the role reversal of EWZ, which now functions as a net giver rather than a receiver of shocks. MCHI is assuming the role of a net receiver instead of a giver.

Figures 2 and 3 present the graphical representations of dynamic connectedness. Adopting a more dynamic analytical framework is crucial, considering the system's dynamic nature. This study finds the temporal variance of TCI and reflects the temporal variability of the studied market role in the network. This study investigates the temporal variation of total connectedness estimations, presenting intertemporal changes in TCI, as shown in Figure 2.

In Figure 3, the bold line between BTC and MCHI signifies substantial spillovers from BTC to MCHI. Additionally, bold lines are observed between VIX and EZA and between BTC and EZA. Figure 3 displays bolder lines than Figure 1, indicating heightened connectedness. Mainly, BTC exhibits robust relationships with MCHI, EZA, INDA, and EZA in the TVP-VAR model, relationships that were absent in the VAR model. Bold lines are evident between VIX, EZA, and ERUS. EWZ, ERUS, VIX, and BTC function as net receivers of shocks, while EZA, INDA, and MCHI serve as net givers of shocks. The dynamic connectedness network in the TVP-VAR model reveals nuanced interactions not captured by the VAR model, providing a clearer understanding of the hedging capabilities in managing dynamic risks.

**Insert Figure 2 Here**

**Insert Figure 3 Here**

**Insert Figure 4 Here**

**Insert Figure 5 Here**

Next, we partition the dataset into two subsets based on the occurrence of the COVID-19 pandemic, one representing the period before and the other after the onset of the pandemic. The nonparametric directional connectivity (NPDC) measure plots from the TVP-VAR connectedness analysis are illustrated in Figures 4 and 5. These figures reveal that EZA consistently acts as a net giver before and after the pandemic. Notably, MCHI functions as a net giver before the COVID-19 pandemic but transforms into a net receiver after the pandemic. Conversely, the VIX shifts from being a net receiver to a net giver post-pandemic.

Table 8 provides insight into the dynamics of a financial market network, where the diagonal elements represent the volatility of each market. The off-diagonal lines encapsulate the interplay between markets, illustrating how each market contributes to the volatility of others. Rows detail the individual variables' contribution to the forecast error variance of a specific variable, while columns depict the impact of specific market types on forecast error variance. Across the entire dataset, the TCI average is 51.89%, signifying that fluctuations within this network can explain over half of the variance in the considered variables. Conversely, approximately 48.11% of the error variance is attributed to other idiosyncratic effects.

**Insert Table 8 Here**

The last row of Table 8 delineates the role of each market. Notably, the South African stock market plays a minor role in transmitting shocks and volatility to other markets. Cryptocurrency, investor sentiment, and the stock markets of Russia and Brazil emerge as net receivers of corresponding shocks, with cryptocurrency holding the pivotal position. This study divides the sample into two subsets based on the onset of the COVID-19 pandemic to uncover temporal disparities. The network exhibits considerable explanatory power before (TCI: 52.27%) and after (TCI: 52.36%) the pandemic; however, substantial shifts occur in net receivers and givers. For instance, the stock markets of Brazil and India, along with investor sentiment, transitioned from net receivers to net givers post-pandemic. Although the Chinese stock market remains a net giver, the magnitude decreases significantly. Cryptocurrency maintains its role as a net receiver, with the net value becoming more negative after the pandemic, reflecting increased sensitivity to shocks (−3.75 to −6.62).

**Insert Figure 6 Here**

**Insert Figure 7 Here**

## 5 Discussion

### 5.1. Impact of cryptocurrency on BRICS stock markets

The results of VAR and TVP-VAR models show that the relationship between cryptocurrency and BRICS stock markets is multifaceted and contingent upon diverse elements such as market sentiments, regulatory frameworks, investor conduct, and overarching economic circumstances. The dynamic nature of the cryptocurrency market introduces an additional layer of intricacy to its interplay with traditional stock markets in the BRICS nations.

Cryptocurrencies, notably Bitcoin, exhibit indications of integration with traditional financial markets. Positive sentiments within cryptocurrency may spill into the BRICS stock markets, particularly when investors perceive digital assets as alternative investment opportunities. Establishing clear and favorable regulatory frameworks for cryptocurrencies enhances the legitimacy of this asset class, attracting increased institutional investor participation. With positive regulatory developments, cryptocurrencies have the potential to flourish in an atmosphere of economic optimism and growth. Positive global economic conditions set a favorable backdrop for the cryptocurrency and stock markets across the BRICS nations. Recognizing these markets' nuanced dynamics and interconnectedness is crucial for understanding their collective impact on the broader financial landscape. Findings from this study show that cryptocurrency is the largest net receiver and is prone to receive shocks from stock markets in the BRICS nations. The stock markets of South Africa and India were verified to be net givers of shocks in both models, and Russia's stock market, cryptocurrency, and investor sentiments were net receivers in both models. Results of the VAR conditional correlation test show that cryptocurrency and the Indian stock market are highly correlated, and the VAR partial correlation results indicate a higher correlation between cryptocurrency and Brazil's stock market. These findings suggest strong interconnectedness and spillover effects between these financial markets.

### 5.2. Impact of investor sentiments on BRICS stock markets

Investor sentiments are pivotal in shaping the dynamics of BRICS stock markets, influencing market movements, investor decisions, and overall economic conditions. The intricate interplay between sentiments and various market factors contributes to the market complexity and dynamics. Favorable global economic conditions and optimistic sentiment in major financial markets act as catalysts for uplifting the BRICS stock markets, whereas adverse global events, economic downturns, or negative sentiment in key international markets may exert downward pressure on BRICS markets. Understanding the intricate relationship between investor sentiments and market dynamics is crucial for comprehending the nuanced behavior of BRICS stock markets.

VIX serves as a metric to gauge investors' sentiment, showing investors' acceptance of risks and benefits in the stock market and their economic behavior. This study found negative correlations between VIX and the BRICS stock markets, surpassing the relationships between cryptocurrency and the BRICS stock markets. The TVP-VAR dynamic conditional connectedness analysis reveals that investor sentiment predominantly acts as a recipient of shocks within the financial network, indicating a reciprocal influence between BRICS stock markets and investor behavior. Our study also shows that the stock markets of South Africa and India denote a higher spillover on investor sentiments than other stock markets, and investor sentiment and South Africa's stock markets are strongly correlated.

This study reveals considerable interconnections between the six financial markets, underscoring the potential risks for investors in these markets and highlighting the increasing interconnectedness between unexpected and highly uncertain events such as the COVID-19 pandemic. Our findings demonstrate that the occurrence of the pandemic affects the entire network. Investors and managers should be aware of the contagion of uncertainty and risk as early warning signs to reconsider investment strategies.

## 6. Conclusion

This study is the first to thoroughly explore the interconnected dynamics and spillover effects between cryptocurrency, investor sentiments, and BRICS stock markets, offering valuable insights for corporations, investors, and regulators concerning systematic risk and investment strategies of BRICS markets. Applying the TVP-VAR approach, this study indicates that among the five BRICS stock markets, China, India, and South Africa are the primary sources of shocks that can subsequently affect the financial network, whereas Brazil and Russia, along with investor sentiment and cryptocurrency, emerge as net receivers of these spillovers. Throughout the observed duration, the benchmarks exhibit high volatility in cryptocurrency, resulting in substantial value erosion, whereas the BRICS stock markets highlight a more consistent performance, as evidenced by the subdued price fluctuations. Nonetheless, the cumulative returns of BRICS fail to make an impressive impact when juxtaposed with major US stock market indices, and over time, they depict a less remarkable performance when capturing capital appreciation and dividends are considered. These findings suggest that diversification across BRICS stock markets and cryptocurrencies may not yield substantially improved investment performance. Stakeholders navigating investment decisions in this multifaceted landscape should consider the intricate relationship between these markets and the associated complexities.

# Tables

**Table 1.** Summary of the literature on stock market connectedness

| Study | Model | Variables | Scale/location | Data Period |
|---|---|---|---|---|
| Loh (2013) | Wavelet coherence method | Stock market returns | 13 Asia-Pacific markets, European and US stock markets | 01/05/2001–03/16/2012 |
| Mensi et al. (2014) | Quantile regression method | Market index | Global vs BRICS | 1997-2013 |
| Khandaker and Islam (2015) | Standard historical volatility model and R-square estimates | Geographical location, size of the equity market and the availability of the stock return | three emerging and three developed stock market | 01/2001–12/2012 |
| Jiang et al. (2017) | VAR model, Granger causality tests | Daily stock market indices at closing time | U.S., UK, Germany, Japan, Hong Kong, and mainland China | 08/09/2007–03/06/2009 |
| Zhang et al. (2020) | GARCH-BEKK model | Daily closing spot price | G20 stock index | 01/02/2006–12/31/2018 |
| Papadamou et al. (2021) | panel data analysis | Daily data on bond and stock returns | Australia (AUS), France (FR), Germany (DEU), Japan (JPN), India (IND), Italy (IT), Netherland (NL), Switzerland (CH), United Kingdom (UK), and United States of America (USA) | 01/02/2020–04/09/2020 |
| Li et al. (2021) | modified BBQ algorithm | Monetary policy cycle, financial cycles, and business cycle | China | 1998–2018 |
| Uddin et al. (2021) | panel-based EGARCH model | Daily price indices, daily trading volumes, and the daily CBOE VIX index | Thirty-four developed and emerging markets | 07/01/2019–08/14/2020 |
| Chiang (2021) | correlation analysis | Twenty-one aggregate market indices | U.S., Canada, Europe, Latin America, and Asia | 01/1998–04/2019 |
| Wen et al. (2021) | TVP-VAR model | Shanghai Composite Index (SHCI) | China | 05/25/2009–06/24/2020 |
| Alexakis et al. (2021) | Spatial Durbin Model (DSDM) | Stock market index returns | Argentina, Brazil, Canada, Mexico, the USA, and other forty countries | 01/02/2020–04/08/2020 |
| Youssef et al. (2021) | TVP-VAR mode | Daily stock indices in China (SSE) France (CAC40), Germany (DAX30), Italy (FTSE MIB), Russia (RTSI), Spain (IBEX), the | China, Italy, France, Germany, Spain, Russia, the US, and the UK | 01/01/2015–05/18/2020 |
| | | UK (FTSE100), and the US (S&P500) | | |
| Rai and Garg (2022) | DCC-GARCH model, BEKK-GARCH | Daily data on stock prices and nominal exchange rates | BRIICS | 01/02/2020–09/15/2020 |
| Moslehpour et al. (2022) | VAR model, DCC-GARCH model | Morgan Stanley Capital International (MSCI) index | United States, Italy, Spain, Germany, China, France, the United Kingdom, Türkiye, Switzerland, Belgium, the Netherlands, Canada, Austria, Vietnam, and South Korea | 01/07/2016–07/01/2020 |

Note: Table 1 summarizes studies on the dynamic connectedness within stock markets.

**Table 2.** Literature summary on cryptocurrency and financial assets

| **Source** | **Variables** | **Method** | **Topic** | **Conclusion** |
|---|---|---|---|---|
| Stavroyiannis and Babalos (2017) | Bitcoin, S&P 500 index | GARCH model | Dynamic properties of the Bitcoin and the US market | Bitcoin is not related to other assets and does not interact with traditional assets |
| Baur et al. (2018) | WinkDex | Regression analysis | Whether Bitcoin is a medium of exchange or an asset | Bitcoin is used for speculative investments and does not correlate with stock price fluctuations |
| Paolo and Iman (2018) | Bitcoin, stocks, exchange rates, gold, etc. | VAR model | The price volatility correlation between Bitcoin and stocks, exchange rates, gold, etc. | The market linkage between Bitcoin and stocks, exchange rates, gold, etc., is low. |
| Feng et al. (2018) | Bitcoin, Ethereum, Ripple, Litecoin, NEM, Dash, and Monero | ARMA-GARCH model | The extreme fluctuation characteristics of Cryptocurrency | The volatility characteristics of Bitcoin are similar to those of commodities and have higher volatility than stocks |
| Kristoufek (2018) | Bitcoin and stocks | Wavelet analysis | The efficiency of two Bitcoin markets and its evolution in time | Bitcoin and stocks have similar volatility characteristics |
| Kurka (2019) | Cryptocurrencies, Foreign exchange, Commodities, Stocks, Bonds | VAR model | Connectedness between the main cryptocurrency and traditional Asset classes | The Bitcoin market can be transmitted to traditional financial markets. |
| Tiwari et al. (2019) | S&P 500 index and six cryptocurrencies | Copula-ADCC-EGARCH model | Correlation between stock and cryptocurrency markets | Digital currencies can serve as assets to hedge against stock market risks. |
| Guesmi et al. (2019) | Bitcoin, stock indices, gold, oil, currencies, and VIX | VARMA (1,1)-DCC-GJR-GARCH model | The conditional cross effects and volatility spillover between Bitcoin and financial indicators | The bitcoin market allows for hedging risky investments for different financial assets. |
| Gil-Alana et al. (2020) | Bitcoin, Ethereum, Ripple, Litecoin, Stellar, and Tether, and a variety of financial assets | Fractional integration techniques | The stochastic properties of six significant cryptocurrencies and their bilateral linkages with six stock market indices | Confirmed the crucial role of digital currencies in investors' diversified choices |
| Shahzad et al. (2020) | Gold, Bitcoin, and the individual stock market indices of the G7 countries | AGDCC-GARCH model | Compare gold and Bitcoin for the G7 stock markets | Gold and Bitcoin have distinct safe haven and hedging characteristics |
| Huynh et al. (2020) | Fourteen cryptocurrencies, gold | Transfer entropy approach | The spillover effects in the cryptocurrency market | Among several types of cryptos, Bitcoin is still the most appropriate instrument for hedging |
| Corbet et al. (2020) | Bitcoin, Chinese stock markets (Shanghai SE, Shenzhen SE) | GARCH model and DCC model | Volatility relationship between the main Chinese stock markets and Bitcoin | The volatility relationship between the main Chinese stock markets and Bitcoin evolved significantly during this period of enormous financial stress. |
| Conlon et al. (2020) | Bitcoin, Ethereum, Tethers, stock indices | Downside risk measurement (value at risk and conditional value at risk) | Are cryptocurrencies a safe haven for equity markets? | Bitcoin and Ethereum are not a safe haven for most international equity markets examined. |
| Hsu et al. (2021) | Bitcoin (BTC), Ethereum (ETH), Ripple (XRP) | Diagonal BEKK model | The risk spillovers of three major cryptocurrencies to ten leading traditional currencies and two gold prices | There are significant co-volatility spillover effects between cryptocurrency and traditional currency or gold markets |
| Uzonwanne (2021) | Bitcoin price and stock index prices for five major stock markets (FTSE 100, S&P 500, CAC 40, DAX 30, and Nikkei 225) | VARMA-AGARCH model | Volatility and return spillovers between stock markets and cryptocurrencies | Verified the presence of returns and volatility, spillovers between bitcoin and five major stock markets |
| Zhang et al. (2021) | Bitcoin, equities, bonds, currencies, and commodities | Expectile VAR | Risk spillover between Bitcoin and conventional financial markets | There is a downside risk spillover between Bitcoin and the four assets |

**Table 3.** Detailed descriptions of the variables

| # | Country | Index/Market | Description | TICKER | Description |
|---|---|---|---|---|---|
| 1 | Brazil | IBOVESPA | The ibovespa Index | EWZ | iShares MSCI Brazil ETF |
| 2 | India | NIFTY 50 | A benchmark Indian stock market index | INDA | iShares MSCI India ETF |
| 3 | South Africa | JSE | FTSE/JSE Top 40 | EZA | iShares MSCI Emerging Markets ETF |
| 4 | China | SSE and SZSE Index | Shanghai and Shenzhen component index | MCHI | iShares MSCI China ETF |
| 5 | Russia | MICEX | MOEX Russia Index | ERUS | iShares MSCI Russia ETF |
| 6 | Cryptocurrency | | Bitcoin | BTC | coinbase |
| 7 | VIX | Investor sentiment | Volatility index to measure volatility in the stock market accompanied by market fear | VIX | Center for Research in Security Prices |

**Table 4.** Descriptive statistics of EWZ, INDA, EZA, MCHI, ERUS and VIX

| Variable | Mean | Median | SD | Skewness | Kurtosis | J-B test | Q2(20) |
|---|---|---|---|---|---|---|---|
| EWZ | 0.000 | 0.0008 | 0.0181 | -1.367*** | 11.252*** | 11928.754*** | 49.578*** |
| INDA | 0.000 | 0.0007 | 0.0143 | -0.980*** | 10.134*** | 9477.431*** | 92.149*** |
| EZA | 0.000 | 0.0004 | 0.0151 | -0.479*** | 2.244*** | 529.679*** | 17.618** |
| MCHI | 0.000 | 0 | 0.0184 | -0.072 | 4.328*** | 1667.963*** | 18.120** |
| ERUS | 0.000 | 0 | 0.0231 | -2.559*** | 23.620*** | 51962.601*** | 70.887*** |
| VIX | -0.001 | -5.18E-05 | 0.0228 | 1.693*** | 49.693*** | 220691.152*** | 108.044*** |
| BTC | 0.004 | 0.0043 | 0.5628 | -0.498*** | 15.062*** | 20268.771*** | 56.899*** |

(. p-value≤0.1, * p-value≤0.05, ** p-value≤0.01, ***, p-value≤0.005). The Ljung-Box test indicates that the null hypothesis of no autocorrelation up to order 20 for the squared standard residuals (Q2(20)) cannot be rejected at the 1% significance level.

**Table 5.** ADF test results for variables and their value at first-order difference

| Ticker | ADF test at first difference (Lag) | p-value |
|---|---|---|
| EWZ | -13.02***(9) | 0.000 |
| INDA | -11.70***(13) | 1.58E-21 |
| EZA | -17.62***(7) | 3.86E-30 |
| MCHI | -27.35***(2) | 0 |
| ERUS | -14.50***(7) | 6.01E-27 |
| VIX | -9.85***(23) | 4.42E-17 |
| BTC | -15.28***(7) | 4.53E-28 |

(. p-value≤0.1, * p-value≤0.05, ** p-value≤0.01, ***, p-value≤0.005)

Note: Table 5 demonstrates that all variables are stationary at the first-order difference across all samples.

**Table 6.** VAR connectedness analysis with time on variables for portfolio connectedness analysis

| | EWZ | INDA | EZA | MCHI | ERUS | VIX | BTC | Receiver |
|---|---|---|---|---|---|---|---|---|
| EWZ | 46.11 | 12.64 | 16.69 | 9.28 | 11.89 | 2.73 | 0.66 | 53.89 |
| INDA | 12.25 | 44.38 | 16.02 | 12.48 | 9.79 | 3.87 | 1.21 | 55.62 |
| EZA | 14.52 | 14.55 | 39.71 | 15.61 | 10.33 | 4.66 | 0.62 | 60.29 |
| MCHI | 9.64 | 13.53 | 18.15 | 46.24 | 7.75 | 3.78 | 0.91 | 53.76 |
| ERUS | 12.87 | 11.71 | 12.86 | 8.08 | 50.07 | 3.59 | 0.82 | 49.93 |
| VIX | 4.20 | 6.34 | 8.16 | 5.69 | 5.05 | 70.52 | 0.03 | 29.48 |
| BTC | 1.19 | 2.37 | 1.41 | 1.81 | 1.14 | 0.04 | 92.04 | 7.96 |
| Giver | 54.67 | 61.15 | 73.29 | 52.94 | 45.96 | 18.67 | 4.26 | 310.93 |
| Inc.Own | 100.77 | 105.53 | 113.00 | 99.18 | 96.02 | 89.19 | 96.30 | TCI |
| NET | 0.77 | 5.53 | 13.00 | -0.82 | -3.98 | -10.81 | -3.70 | 51.82 |
| NPT | 4 | 5 | 6 | 3 | 2 | 1 | 0 | |

Givers: EZA, INDA, EWZ

Receivers: MCHI, ERUS, VIX, BTC

**Table 7**. TVP-VAR connectedness analysis on variables

| | EWZ | INDA | EZA | MCHI | ERUS | VIX | BTC | Receiver |
|---|---|---|---|---|---|---|---|---|
| EWZ | 44.80 | 10.11 | 14.65 | 9.97 | 10.06 | 7.97 | 2.44 | 55.20 |
| INDA | 9.79 | 42.76 | 13.99 | 12.08 | 8.42 | 10.05 | 2.91 | 57.24 |
| EZA | 12.52 | 12.51 | 37.14 | 15.49 | 10.16 | 9.03 | 3.15 | 62.86 |
| MCHI | 9.25 | 11.72 | 16.76 | 41.05 | 7.96 | 10.03 | 3.23 | 58.95 |
| ERUS | 10.27 | 9.13 | 12.00 | 9.11 | 48.10 | 8.75 | 2.64 | 51.90 |
| VIX | 8.64 | 10.87 | 10.97 | 11.53 | 8.13 | 46.35 | 3.51 | 53.65 |
| BTC | 3.24 | 4.22 | 4.44 | 4.05 | 3.13 | 4.34 | 76.59 | 23.41 |
| Giver | 53.71 | 58.55 | 72.81 | 62.24 | 47.86 | 50.16 | 17.87 | 363.20 |
| Inc.Own | 98.51 | 101.31 | 109.95 | 103.29 | 95.96 | 96.52 | 94.46 | TCI |
| NET | -1.49 | 1.31 | 9.95 | 3.29 | -4.04 | -3.48 | -5.54 | 51.89 |
| NPT | 3 | 4 | 6 | 5 | 1 | 2 | 0 | |

Givers: EZA, INDA, MCHI

Receivers: EWZ, ERUS, VIX, BTC

Table 8. TVP-VAR connectedness analysis on variables before and after COVID-19

| | Before Covid-19 pandemic | | | | | | | | After Covid-19 pandemic | | | | | | | |
|---|---|---|---|---|---|---|---|---|---|---|---|---|---|---|---|---|
| | EWZ | INDA | EZA | MCHI | ERUS | VIX | BTC | Receiver | EWZ | INDA | EZA | MCHI | ERUS | VIX | BTC | Receiver |
| EWZ | 43.71 | 9.4 | 14.91 | 10.97 | 11.94 | 7.70 | 1.37 | 56.29 | 44.66 | 10.42 | 12.99 | 8.82 | 8.92 | 10.32 | 3.87 | 55.34 |
| INDA | 9.24 | 43.49 | 13.41 | 14.15 | 7.97 | 9.86 | 1.88 | 56.51 | 9.66 | 43.38 | 12.51 | 9.48 | 8.38 | 12.30 | 4.28 | 56.62 |
| EZA | 12.50 | 11.77 | 36.46 | 15.75 | 11.51 | 10.20 | 1.81 | 63.54 | 11.52 | 11.95 | 37.83 | 14.60 | 9.29 | 10.20 | 4.61 | 62.17 |
| MCHI | 9.14 | 12.59 | 15.93 | 37.95 | 9.57 | 12.79 | 2.03 | 62.05 | 8.22 | 9.64 | 16.20 | 43.84 | 6.83 | 10.57 | 4.71 | 56.16 |
| ERUS | 12.27 | 8.6 | 13.69 | 11.51 | 43.39 | 8.85 | 1.69 | 56.61 | 8.11 | 9.19 | 9.35 | 7.20 | 53.52 | 8.93 | 3.70 | 46.48 |
| VIX | 7.82 | 10.71 | 11.52 | 14.86 | 8.80 | 43.97 | 2.32 | 56.03 | 8.78 | 12.51 | 10.58 | 10.90 | 7.75 | 43.76 | 5.73 | 56.24 |
| BTC | 1.81 | 2.97 | 2.05 | 2.55 | 2.34 | 3.14 | 85.15 | 14.85 | 4.16 | 5.46 | 5.84 | 5.86 | 4.66 | 7.54 | 66.48 | 33.52 |
| Giver | 52.77 | 56.04 | 71.52 | 69.78 | 52.12 | 52.55 | 11.10 | 365.87<br>TCI | 50.45 | 59.17 | 67.47 | 56.87 | 45.83 | 59.86 | 26.90 | 366.54 |
| NET | -3.52 | -0.47 | 7.98 | 7.74 | -4.49 | -3.49 | -3.75 | 52.27 | -4.89 | 2.55 | 5.29 | 0.70 | -0.66 | 3.62 | -6.62 | 52.36 |
| NPT | 3 | 4 | 6 | 5 | 1 | 2 | 0 | | 1 | 5 | 6 | 4 | 2 | 3 | 0 | |

Note: the diagonal elements represent the volatility of each market.

**Figures**

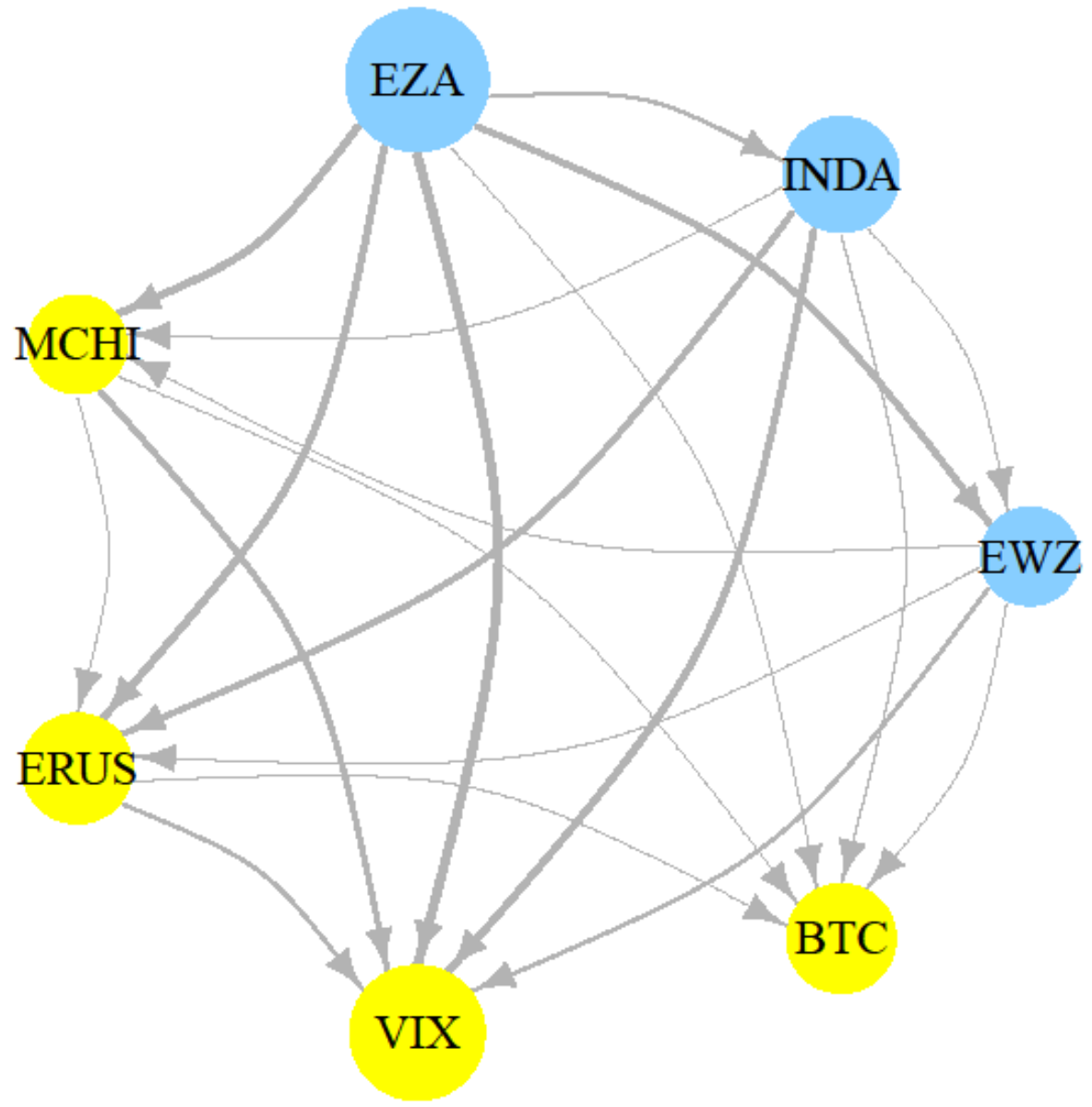

Figure 1. NPDC measure plot of VAR connectedness with time

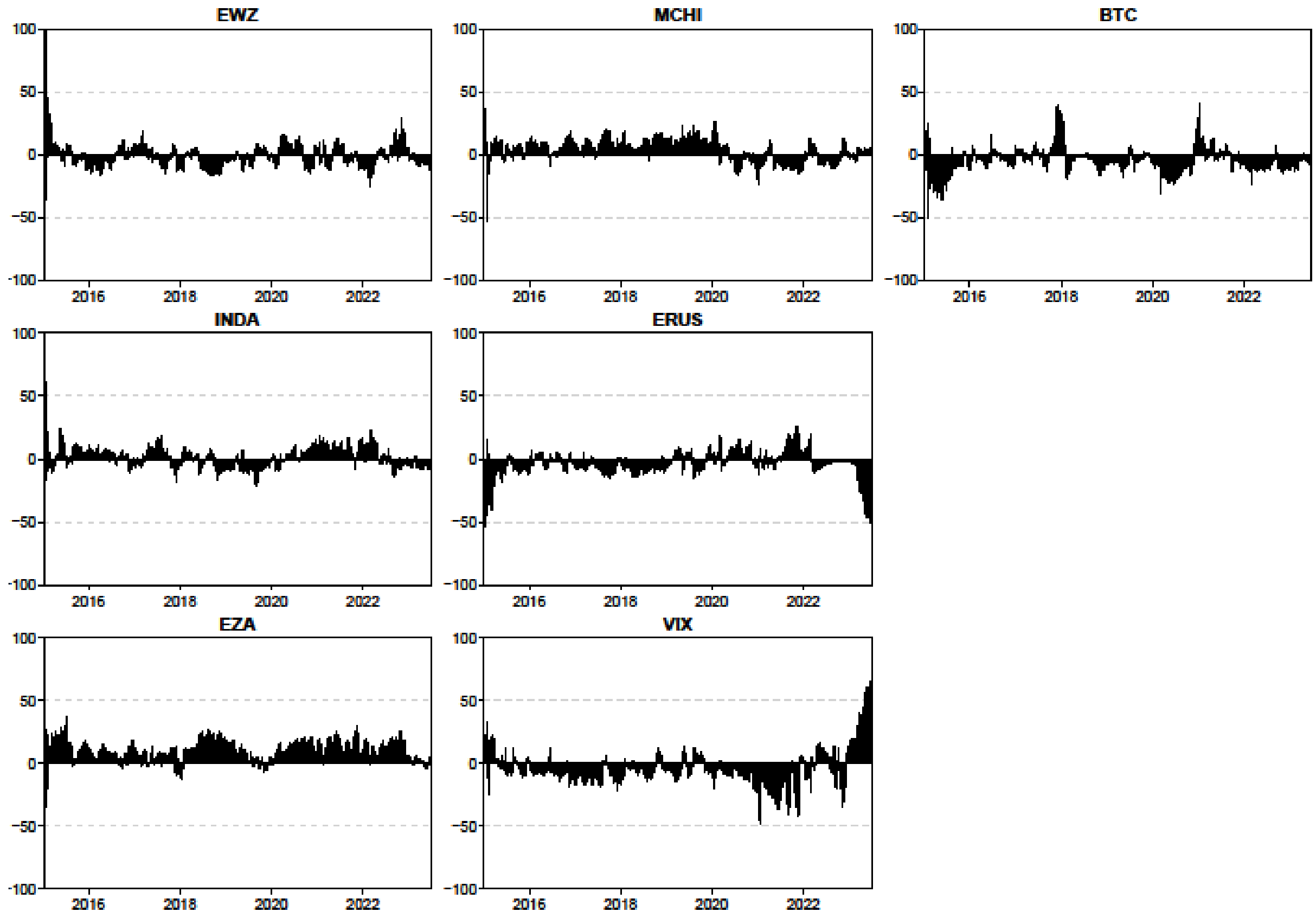

Figure 2. Net value plot on all periods

Note: Figure 2 presents intertemporal changes in TCI.

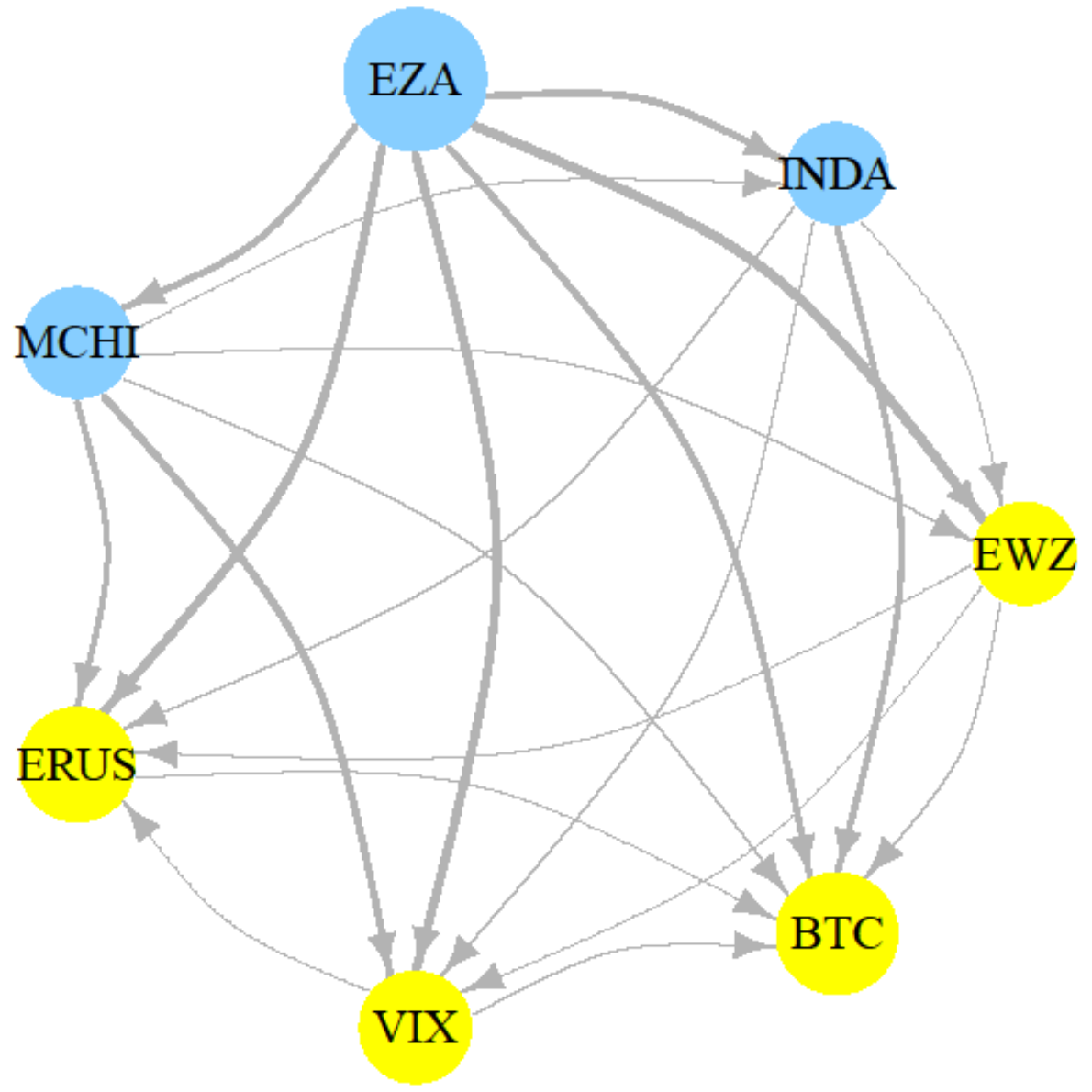

Figure 3. NPDC measure plot of the TVP-VAR connectedness analysis

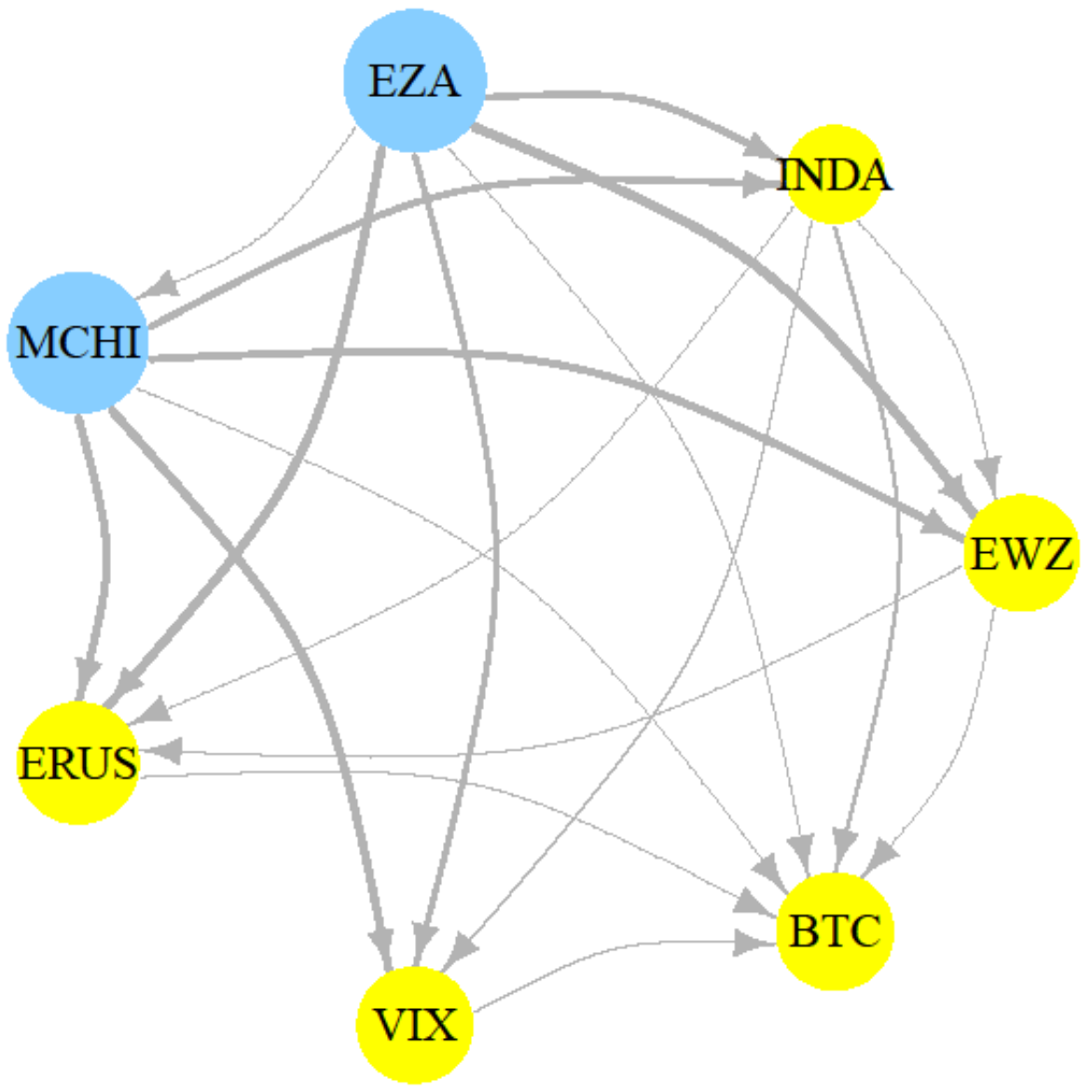

Figure 4. NPDC measure plot of the TVP-VAR connectedness analysis before COVID-19

Note: nodes in blue act as net givers, and nodes in yellow are net receivers of shocks in the financial network. Arrows in the graph indicate greater directional connectedness concerning net connectedness. Node sizes in the graph reflect the weighted average net total directional connectedness in the financial network. Bold lines in the graph denote a higher spillover than fine lines between variables.

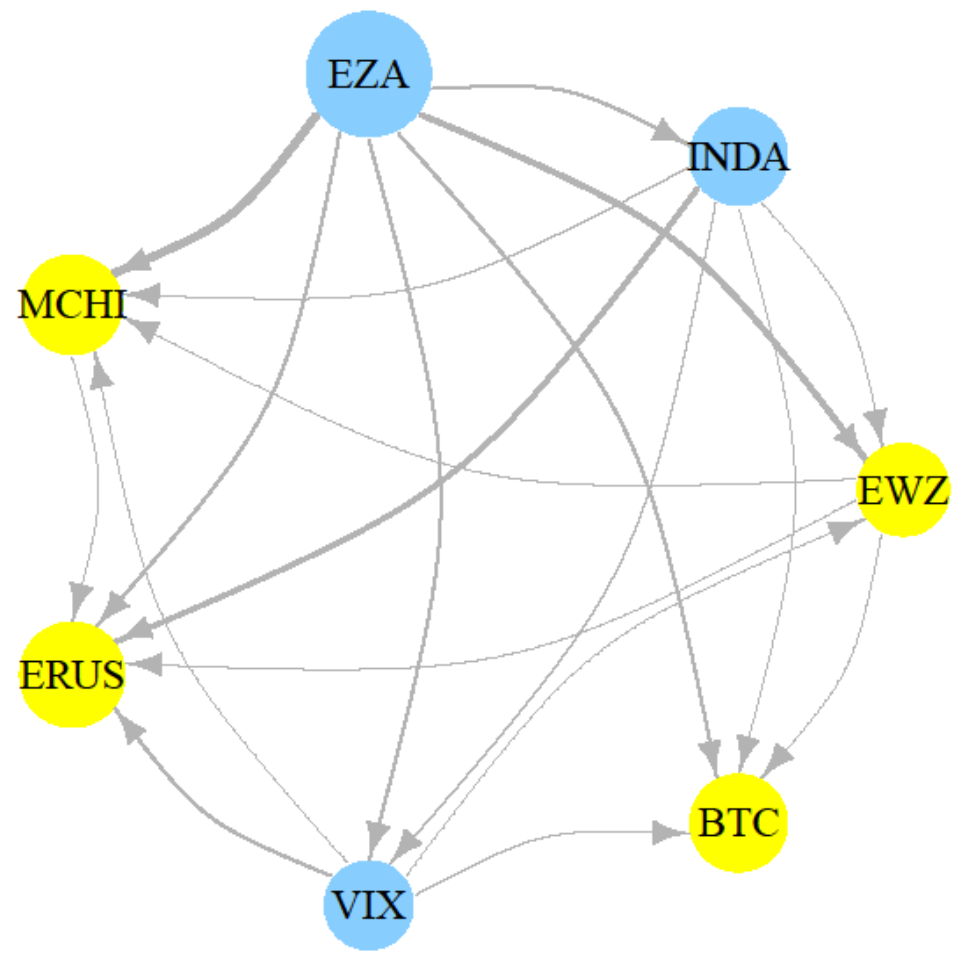

Figure 5. NPDC measure plot of the TVP-VAR connectedness analysis after COVID-19

Note: nodes in blue act as net givers, and nodes in yellow are net receivers of shocks in the financial network. Arrows in the graph indicate greater directional connectedness concerning net connectedness. Node sizes in the graph reflect the weighted average net total directional connectedness in the financial network. Bold lines in the graph denote a higher spillover than fine lines between variables.

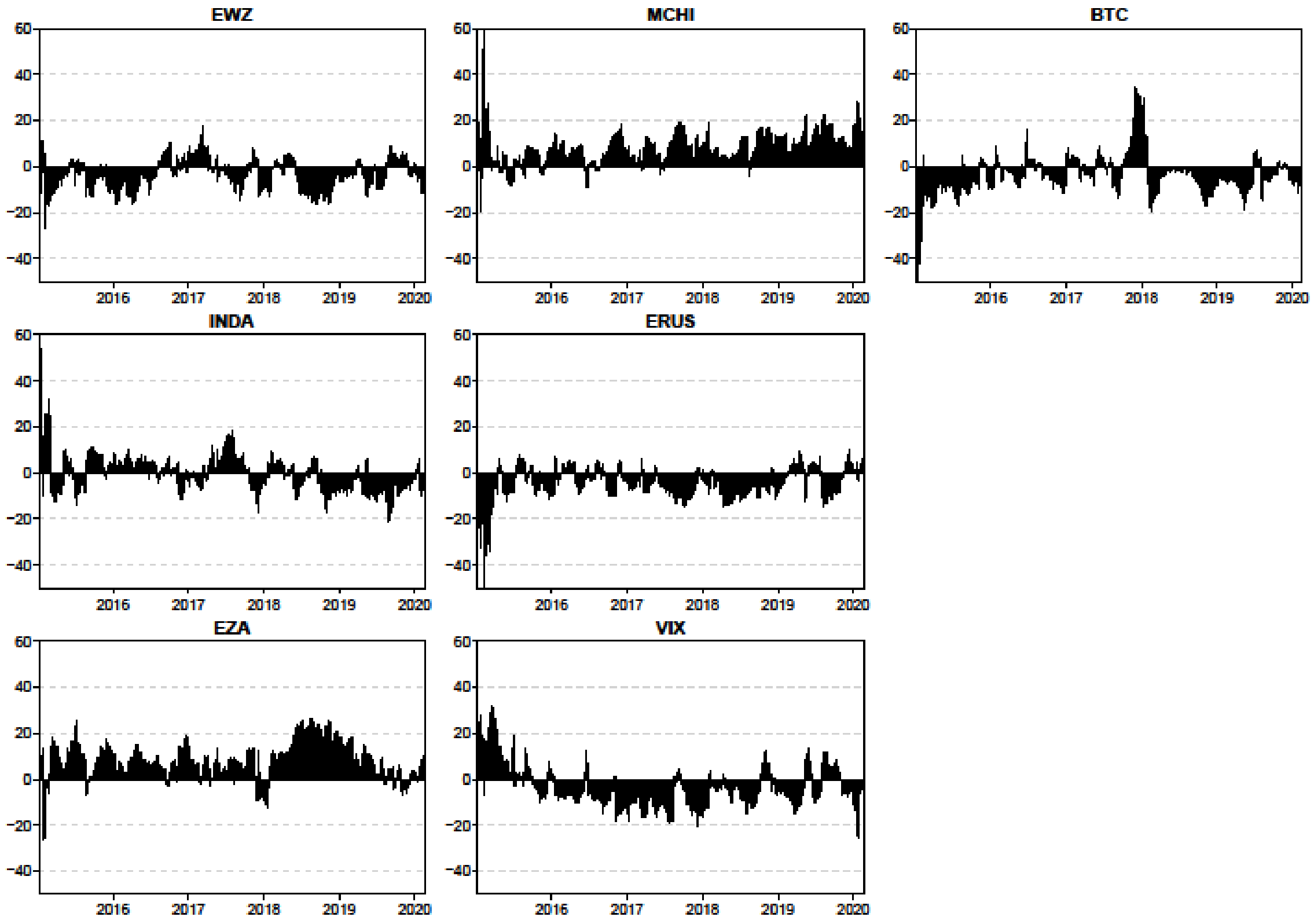

Figure 6. Net value before COVID-19

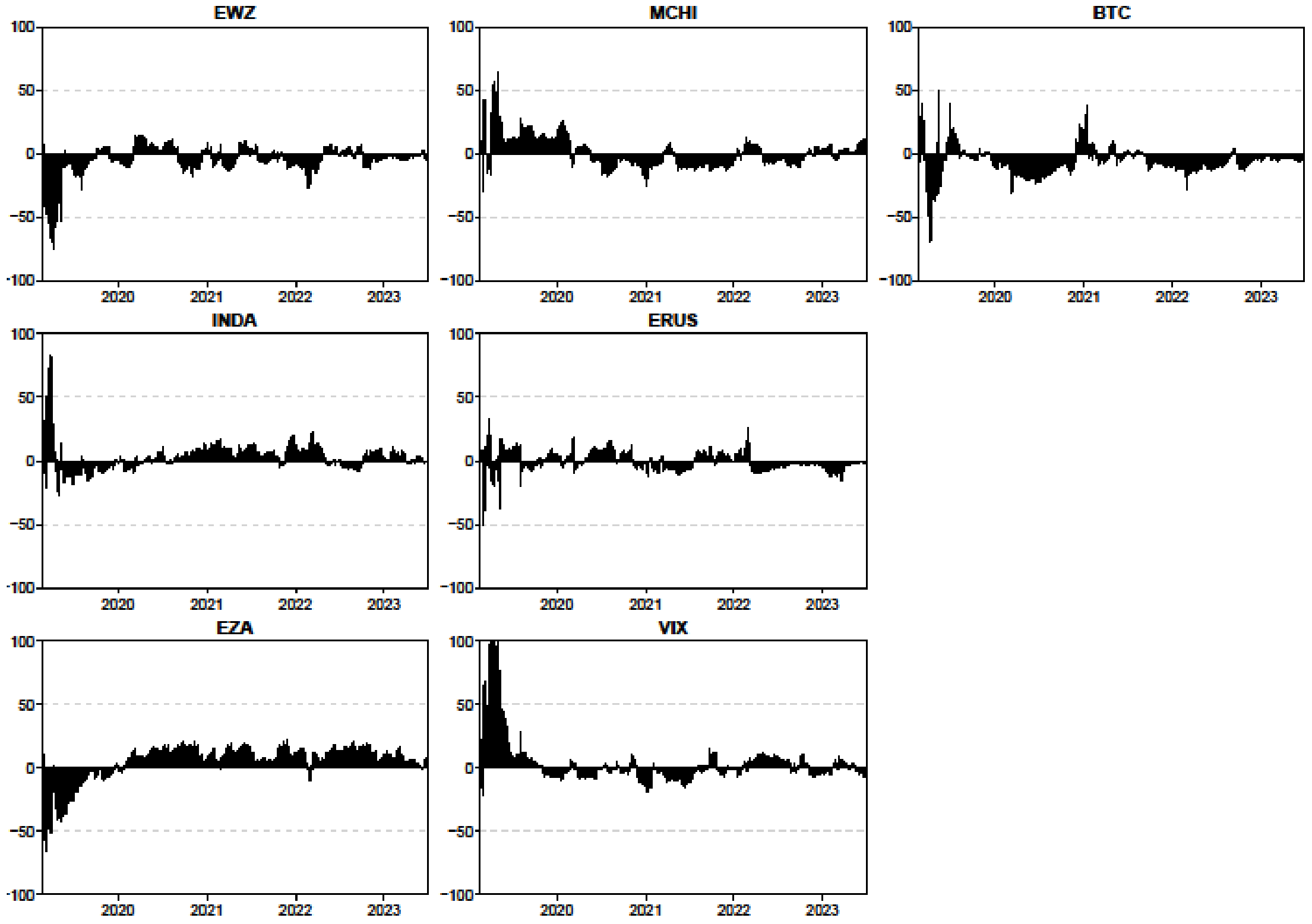

Figure 7. Net value after COVID-1